# Split Radiance Cascades: Real-Time Global Illumination via Sparse Radiance Probes

ROULI FREEMAN, University of Oxford, United Kingdom
ALEXANDER SANNIKOV, Grinding Gear Games, New Zealand
JOE PITT-FRANCIS, University of Oxford, United Kingdom

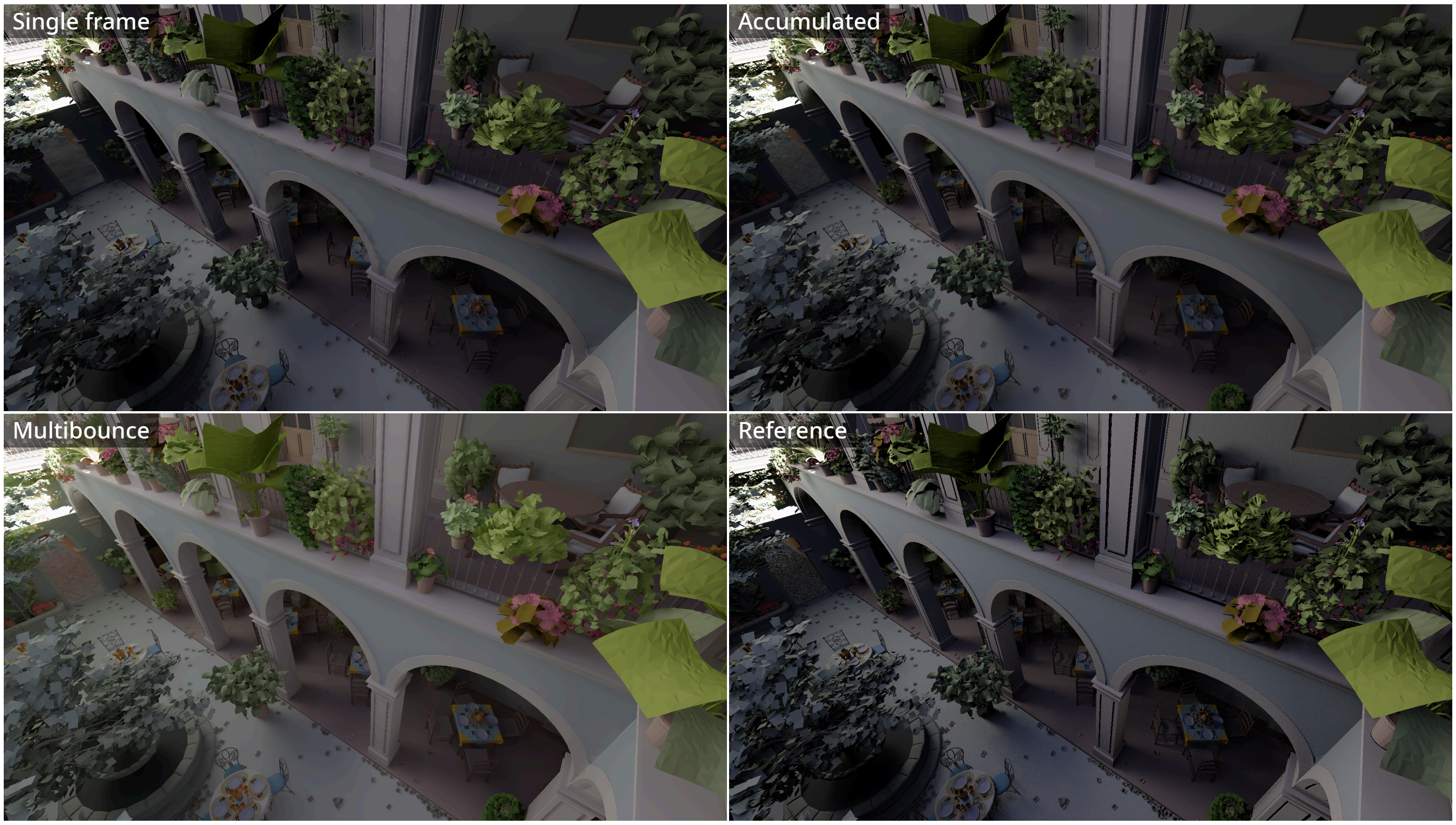

Fig. 1. Showcase of Split Radiance Cascades (Split RC) calculating a single bounce of indirect illumination for the San Miguel scene. “Single frame” shows the result of Split RC after a single frame of calculation; “Accumulated” is a result after 10 frames of temporal accumulation; “Multibounce” adds a secondary offscreen cache for multiple bounces of indirect lighting; “Reference” shows a path-traced reference. Runtime for Split RC in “Single frame” is 13.2ms, of which 6.1ms consists of 1spp raytracing.

Radiance probe methods are a popular and well-tested approach for approximating diffuse global illumination for real-time graphics, but they commonly suffer from a lack of detail due to the large spacing between probes. *Radiance Cascades* (RC) fixes this by increasing spatial resolution and reducing angular resolution for light and occlusion from closer objects, which allows it to provide details at all scales without noise or aliasing. However, leading implementations of RC either run in 2D or screenspace, due to the prohibitive costs of storing high-detail volumetric radiance information.

In this work, we adapt Radiance Cascades for accurate real-time 3D diffuse global illumination using a sparse hashmap to store world-space probes. We introduce *ray splitting*, a method for calculating radiance intervals used in RC by tracing rays from visible surfaces and calculating their contribution to cascades based on their hit distance. We evaluate our algorithm, *Split Radiance Cascades*, on a variety of scenes, and demonstrate that it can provide high-quality indirect illumination in both single-frame and temporally accumulated contexts.

## 1 Introduction

Computing accurate global illumination (GI) in real-time is one of the greatest problems of computer graphics. For a long time, games have had to rely on precomputed lighting approximations to achieve high visual quality, which makes dynamic changes to the scene difficult to depict realistically. The primary problem with GI is that all surfaces effectively are emissive due to the multiple bounces of light, which results in quadratic complexity from interactions between all pairs of surfaces visible to each other.

This problem is commonly split into separate steps which solve specular/glossy and diffuse illumination and are combined to produce the full result. Specular effects are (usually) highly directional and so can be handled efficiently by sampling the BRDF; in contrast, diffuse effects require gathering light from all directions in the visible hemisphere, however they can be temporally accumulated easily due to being view-independent. We will only consider diffuse effects in this work.

Irradiance probes [Greger et al. 1998] have frequently been used as a solution to this problem. While their use in game engines has historically involved precomputation, methods such as *Dynamic Diffuse Global Illumination* (DDGI) [Majercik et al. 2019] have been introduced which calculate irradiance at probes dynamically using raytracing. Additionally, by storing information for all surfaces near the camera, DDGI can efficiently approximate multibounce diffuse effects by feeding bounce information from previous frames into later invocations of the algorithm temporally. However, to avoid noise, (ir)radiance probe methods commonly have a low spatial resolution and so cannot resolve small-scale effects such as ambient occlusion.

*Radiance Cascades* (RC) [Sannikov 2023] are a variant of radiance probes which introduces multiple co-located cascades of probes, which each provides radiance and transmittance information from a different distance interval. By storing cascades which track faraway objects with a lower spatial resolution but higher angular resolution, and vice versa for nearby objects, RC can minimize error for effects caused by occluders at all distances. RC delivers state of the art results in 2D, as can be seen in games such as *Core Keeper* and *Kyora*, and has been demonstrated as a screenspace GI algorithm to great effect in *Path of Exile 2*. However, it has not been explored much for full 3D diffuse illumination including offscreen lights and multiple bounces.

In this work, we first adapt Radiance Cascades for 3D using a sparse hashmap to store probes near surfaces, and sparse trilinear interpolation to calculate irradiance per-pixel. We use multiple levels of detail similarly to cascaded shadow maps [Dimitrov 2007] to ensure constant visual size of probes onscreen. We then develop *ray splitting*, a method of converting full rays into ray intervals for use in RC, and adjust the algorithm to randomly trace rays from visible surfaces and use the intervals to populate the RC data structure in a similar manner to SHaRC [NVIDIA 2024] or hashed path space filtering [Binder et al. 2021]. This avoids having to calculate intervals for each cascade separately, which reduces raytracing cost by the total number of cascades. As this approach requires an even distribution of rays, we demonstrate a method to choose ray directions which ensures uniform coverage across all cascades. We evaluate our full algorithm, *Split Radiance Cascades* (Split RC), and demonstrate that it is capable of providing accurate one-bounce results in real-time with only a single frame of calculation while also being capable of temporal accumulation for increased quality and further bounces.

## 2 Related Work

Work on global illumination spans several decades. Ritschel et al. [2012] provides a summary of approaches up to 2012. Recently, methods utilizing monte-carlo raytracing and spatiotemporal reuse [Ouyang et al. 2021; Lin et al. 2022] have gained popularity as they reliably converge to the reference solution; however for real-time applications, they rely heavily on temporal accumulation, per-pixel reprojection, and denoisers. As managing the resulting tradeoffs is beyond the scope of our paper, we will primarily focus on other algorithms that bear more similarity to our approach.

*(Ir)radiance Probes.* Irradiance probe volumes have been used in game engines frequently as a precomputed approximation of diffuse global illumination since their introduction in Greger et al. [1998]. Storing directional radiance (such as via spherical harmonics) instead of irradiance to render specular effects has also been explored [Krivanek et al. 2005]. More recently, these techniques have been adapted for dynamic environments and real-time calculation. DDGI [Majercik et al. 2019; Majercik et al. 2021] utilizes raytracing in order to update probes in real-time, and stores additional depth moments to prevent light leaking by modifying the interpolation function. Due to probe volumes scaling cubically with resolution, recent techniques have focused on storing radiance only near surfaces, either via surfels [Halen et al. 2021], sparse hashmaps [NVIDIA 2024], or texture mapping [Tatzgern et al. 2024]. Techniques such as *Lumen* [Wright et al. 2022] and *GI-1.0* [Boissé et al. 2023] use an additional screenspace radiance cache for first-bounce lighting to prevent the detail loss from the low spatial resolution of most world-space probes.

*Radiance Cascades.* The original Radiance Cascades paper [Sannikov 2023] provides formulations of RC for 2D, pure screenspace, and screenspace with world-space raytracing. Osborne and Sannikov [2024] introduces the *bilinear fix* method which removes ringing artifacts in 2D implementations. *Holographic Radiance Cascades* [Freeman et al. 2025] is a variant of RC which utilizes anisotropic probe layouts to eliminate ray gaps larger than the base cascade probe spacing, which results in every cascade having the same accuracy; however, achieving the same effect in 3D is problematic as it would require $O(x^4)$ memory for a scene $x$ times the base cascade probe spacing in size.

There have been three fully-3D variants of RC for global illumination. Mathis [2024] stores probes in 2D using texture mapping, which allows the problem to be considered identically to 2D RC. Surfel RC [Coppen 2025] uses surfels to initialize probes on surfaces, however interpolation artifacts are noticeable (although the painterly look may be a feature for some aesthetics). *Love and Deepspace* utilizes Radiance Cascades in 3D [Ruan 2026] to increase the efficiency of standard irradiance probes for use on mobile devices.

*Path Space Filtering.* Binder et al. [2021] traces rays from the camera, and then averages similar rays together using a world-space voxel hashmap and atomic operations. To avoid discretization artifacts, it jitters query and inserted ray positions. SHaRC [NVIDIA 2024] utilizes a similar hashmap structure with multiple levels of detail and no jittering; it is targeted towards second-bounce effects, so artifacts are much less noticeable as the cache is not directly displayed on-screen.

Table 1. Terminology used throughout this paper.

| Symbol | Definition |
|---|---|
| $L(\mathbf{x}, \omega_o)$ | Radiance leaving a point $\mathbf{x}$ in direction $\omega_o$ |
| $L_e(\mathbf{x}, \omega_o)$ | Emission from a point $\mathbf{x}$ in direction $\omega_o$ |
| $L_i(\mathbf{x}, \omega_i)$ | Incident radiance at a point $\mathbf{x}$ from direction $\omega_i$ |
| $\vec{\boldsymbol{n}}(\mathbf{x})$ | Normal vector at a point $\mathbf{x}$ |
| $\rho(\mathbf{x}, \omega_o, \omega_i)$ | Bidirectional Reflectance Distribution Function (BRDF) |
| $\Omega(\mathbf{x})$ | Hemisphere of directions facing away from the surface at point $\mathbf{x}$ |
| $X^I$ | Set of radiance probe positions |
| $\Omega^I$ | Set of ray directions per probe |
| $I(\mathbf{p}, \omega_p)$ | Computed incident radiance for probe $\mathbf{p} \in X^I$ and direction $\omega_p \in \Omega^I$ |
| $\Delta^s$ | Spacing between radiance probes |
| $\Delta^\omega$ | Average spacing between ray directions at some distance |
| $N$ | Total number of cascades |
| $X_n^I, \Omega_n^I, \Delta_n^s, \Delta_n^\omega$ | Values of $X^I, \Omega^I, \Delta^s, \Delta^\omega$ for cascade $n$ |
| $t_n$ | Ray cutoff length for cascade $n$ |
| $K$ | Ray direction count ($\lvert\Omega_n^I\rvert$) increase per cascade (aka. branching factor) |
| $l$ | Ray length ($t_n$) increase per cascade (aka. length scaling coefficient) |
| $J_n(\mathbf{p}, \omega_p)$ | Computed radiance along the interval $t_{n-1} < t \leq t_n$ |
| $\beta_n(\mathbf{p}, \omega_p)$ | Computed transmittance across the interval $t_{n-1} < t \leq t_n$ |
| $I_n(\mathbf{p}, \omega_p)$ | Average radiance in the cone around direction $\omega_p$, for distances $t > t_{n-1}$ |
| $m_n(\omega)$ | The nearest direction in $\Omega_n^I$ to some direction $\omega$ |

## 3 Background

One of the main goals of graphics is approximating the rendering equation [Kajiya 1986] for radiance reflecting off surfaces:

$$L(\mathbf{x}, \omega_o) = L_e(\mathbf{x}, \omega_o) + \int_{\Omega(\mathbf{x})} \rho(\mathbf{x}, \omega_o, \omega_i) L_i(\mathbf{x}, \omega_i)\,(\omega_i \cdot \vec{\boldsymbol{n}}(\mathbf{x}))\,d\omega_i \tag{1}$$

This defines the radiance $L$ at position $\mathbf{x}$ in the direction $\omega_o$ as a function of the emission $L_e$, incident radiance $L_i$, the normal $\vec{\boldsymbol{n}}(\mathbf{x})$ at the position, and the BRDF $\rho$, integrated over the hemisphere $\Omega(\mathbf{x})$ of directions facing away from the surface at that point. The incident radiance $L_i$ is then a function of $L$ at another location; specifically, we can write it as $L_i(\mathbf{x}, \omega) = L(\mathbf{x} + t \cdot \omega, -\omega)$, where $t$ is the distance to the first collision for a ray starting at $\mathbf{x}$ in the direction $\omega$.

As a consequence of this formulation, radiance is defined in terms of itself. In order to avoid the recursive dependence, we split radiance into multiple bounces, where each bounce uses the previous bounce as the incident radiance. The solution to the rendering equation can be calculated by repeating this process until convergence.

For the purposes of our algorithm, we will focus on calculating a single bounce of radiance. Our algorithm supports gathering diffuse illumination at arbitrary surface locations rather than purely on-screen like some final gathering methods, so we can easily rerun it multiple times or feed the output into itself temporally to get multiple bounces.

### 3.1 Radiance Probes

Radiance probe methods approximate $L$ by discretizing the sample space over which the incident radiance $L_i$ is calculated. This is important as computing $L_i$ requires a ray trace (or an equivalent operation), which is expensive and can make up a large portion of the runtime.

Formally, we calculate $I(\mathbf{p}, \omega_p) = L_i(\mathbf{p}, \omega_p)$ for all probe positions $\mathbf{p}$ in some finite set $X^I$ and ray directions $\omega_p$ in a set of evenly spaced unit vectors $\Omega^I$. Then, we can replace the hemisphere integral in Eq. 1 with a sum over the discretized angles:

$$\frac{2\pi}{\lvert\Omega^I \cap \Omega(\mathbf{x})\rvert} \sum_{\omega_p \in \Omega^I \cap \Omega(\mathbf{x})} \rho(\mathbf{x}, \omega_o, \omega_p) \cdot \mathbf{Interp}(I)(\mathbf{x}, \omega_p)\,(\omega_p \cdot \vec{\boldsymbol{n}}(\mathbf{x})) \tag{2}$$

The position at which we calculate radiance is unlikely to coincide with one of the probes, so we extend the function $I$ to all surface points using some interpolation method $\mathbf{Interp}$. For example, a simple method would be trilinear interpolation, if we have probes spaced in a uniform 3D grid: $X^I \subset \{\Delta^s v \mid v \in \mathbb{Z}^3\}$, where $\Delta^s$ is the average probe spacing. However, trilinear results in light leaking artifacts when there are surfaces thinner than the probe spacing. One way of fixing this would be to only interpolate probes which are visible from the sample position [Majercik et al. 2019].

Then, if we need to calculate radiance within a volume of size $V$, the total amount of work necessary is approximately $O\left(\lvert\Omega^I\rvert\ V/(\Delta^s)^3\right)$, assuming that the greatest cost is in the ray-tracing to compute $I$.

There are two main unavoidable forms of artifacts for radiance probe techniques: angular and spatial errors. Consider a probe spaced a distance $t$ away from an emitter. Then, near the emitter, we have average spacing between rays,

$$\Delta^\omega = t\sqrt{\frac{4\pi}{\lvert\Omega^I\rvert}} \tag{3}$$

so if the size of the emitter is less than this, we are likely to miss it. This type of error results in streaks of light, as shown in Fig. 2, or noise if the ray directions vary between probes.

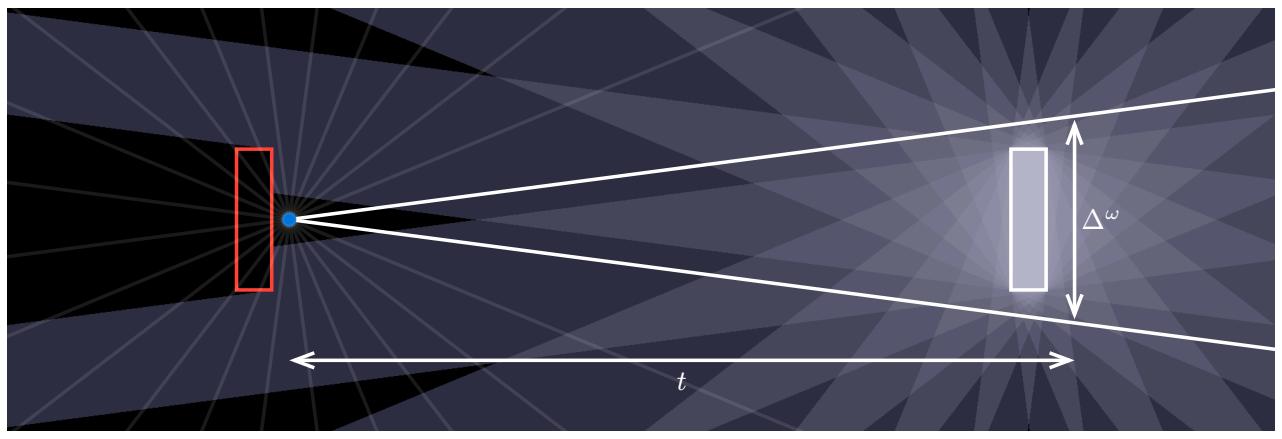

Fig. 2. A scene with an occluder and an emitter, with lighting calculated using a small number of directions. The angular error is larger than the emitter, so the blue probe receives no radiance.

Similarly, consider an occluder near the probe. If the occluder is smaller than the probe spacing $\Delta^s$, we have no guarantee that any of the rays towards the emitter will first hit the occluder, which results in light leaking and missed shadows for positions interpolated from the probes, as shown in Fig. 3.

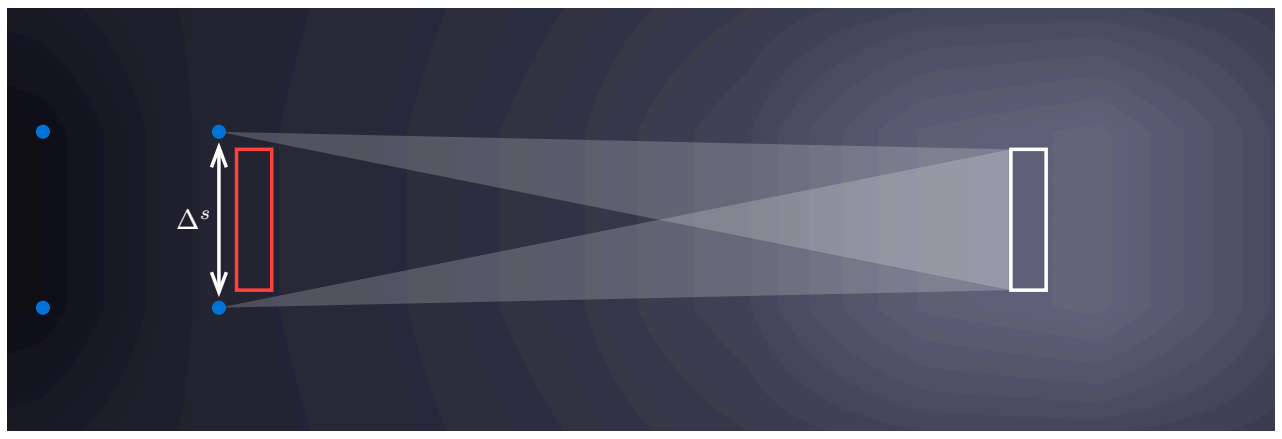

Fig. 3. The same scene as the previous diagram, but with a large probe spacing and high angular resolution instead. The occluder is not picked up as it is between probes, resulting in light leaking.

Therefore, to reduce the resolvable object size, we seek to minimize $\max(\Delta^s, \Delta^\omega)$. Note that depending on the specific circumstances (such as what denoiser is being used), the noise from large angular error could appear worse than the blurring and leaking from spatial error or vice versa, so the specific scaling of each may vary; but for our algorithm this heuristic has worked well. As we have a finite computational budget, we can achieve this best by setting $\Delta^s = \Delta^\omega$. Unfortunately, $\Delta^\omega$ is not constant; it scales linearly with the distance from the probe (see Eq. 3), so as the distance to emitters increases, $|\Omega^I|$ also has to increase, and due to the finite computational budget, $\Delta^s$ as well.

Normal radiance probe techniques have to consider long-distance light sources in order to do global illumination; as a consequence, they usually have a low spatial resolution to avoid noise, and cannot provide high-frequency details such as shadowing within areas smaller than the spacing.

This is not the case for all algorithms. Consider ambient occlusion (AO) techniques: Instead of concerning itself with distant sources, AO assumes that radiance from beyond a certain radius is uniform, and so only calculates visibility within a local area. This can be done by tracing a few short rays (as due to the length, they have a low angular error), and so it can be calculated per-pixel in a reasonable amount of time.

In real-time graphics, these two techniques are commonly combined to approximate global illumination, as AO can add back some of the high-frequency detail that irradiance probes miss. However, there is a gap between the detail levels handled by both techniques, which makes the combined algorithm unable to handle effects such as soft penumbras.

But there is no reason we need to stick with only two levels of detail: if we can accurately combine information from probe grids with exponentially increasing cutoff lengths, we could get something that provides close-to-optimal detail for light sources and occluders at any distance. This is the core idea behind Radiance Cascades.

### 3.2 Radiance Cascades

Radiance Cascades [Sannikov 2023; Osborne and Sannikov 2024] tracks multiple overlapping cascades of radiance probes: Starting from a base grid with spacing $\Delta^s_0$, set of directions $\Omega^I_0$, and ray cutoff length $t_0$, each higher cascade exponentially increases these values. Specifically, we have for the $n$th cascade,

$$\begin{aligned} \Delta^s_n &= \Delta^s_0 \cdot 2^n \\ |\Omega^I_n| &= |\Omega^I_0| \cdot K^n \\ t_n &= t_0 \cdot l^n \end{aligned} \tag{4}$$

for some *branching factor* $K$ and *length scaling coefficient* $l$. We need to be slightly careful about the probe positions: as we will be using trilinear interpolation, we offset each cascade to evenly distribute error: $X^I_n \subset \left\{\Delta^s_n\left(v + \frac{1}{2}\right) \mid v \in \mathbb{Z}^3\right\}$. We will also want the same property for the directions. An example configuration for 2D is shown in Fig. 4.

The angular error at the end of rays is thus:

$$\Delta^\omega_n = t_n \sqrt{\frac{4\pi}{|\Omega^I_n|}} = \left(\frac{l}{\sqrt{K}}\right)^n \Delta^\omega_0 \tag{5}$$

So, setting $K = 4$ and $l = 4$ results in $\Delta^\omega$ scaling as $2^n$, and then adjusting $t_0$ allows us to achieve $\Delta^\omega_n = \Delta^s_n$ for all cascades. Prior to this work, most RC implementations have actually aimed for constant $\Delta^\omega$ across cascades by setting $l = 2$, however in our experiments aiming for equal angular and spatial error using $l = 4$ reduces artifacts significantly, as we demonstrate in Appendix A.

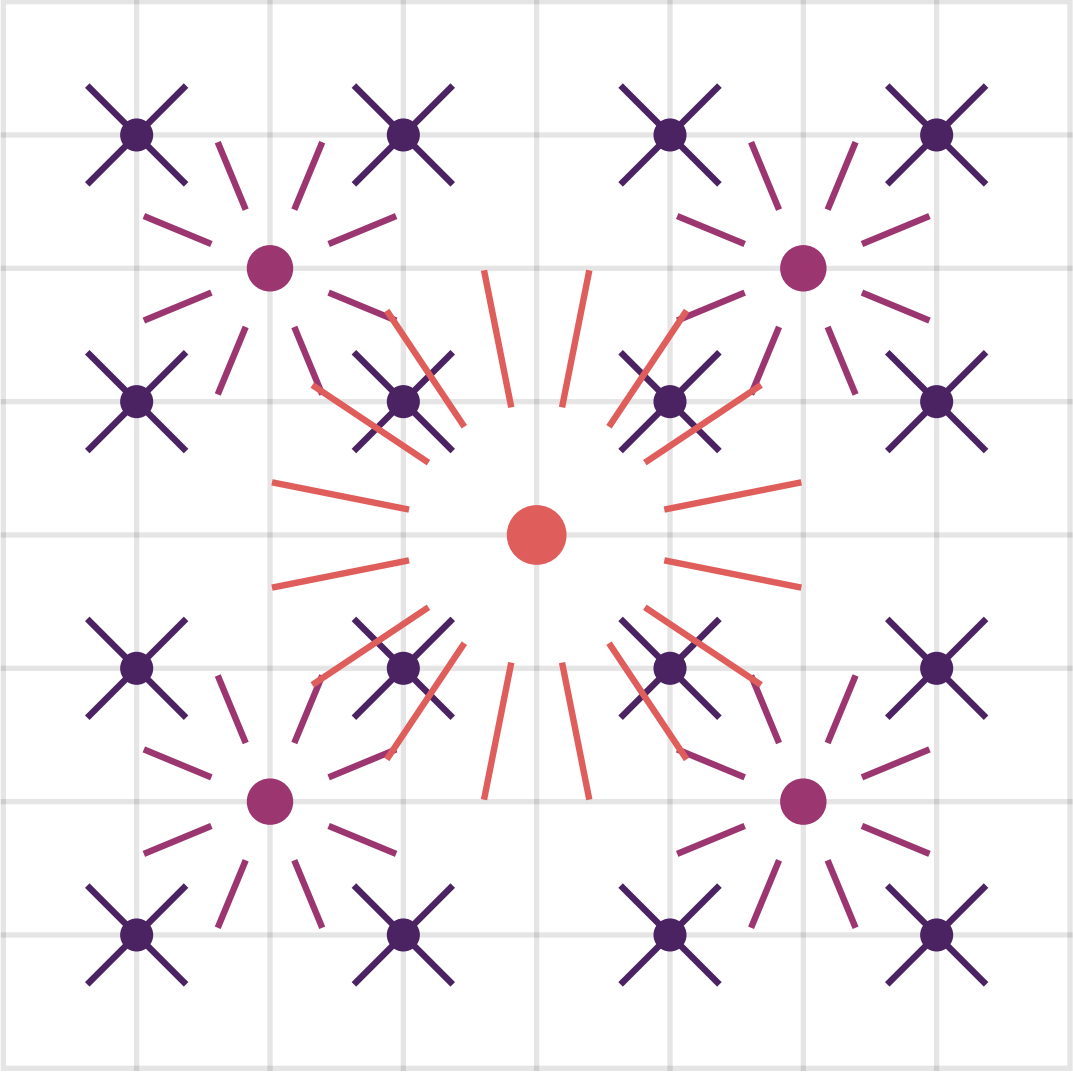

Fig. 4. The first three cascades of probes and the rays computed in them for 2D Radiance Cascades, using settings $K = l = 2$, $|\Omega^I_0| = 4$. Probe positions are offset to reduce aliasing.

Now, consider a ray traced from some position $\mathbf{x}$ in direction $\omega$ to calculate the incident radiance $L_i(\mathbf{x}, \omega)$. We can split this ray into multiple segments: if the first occluder is between distance 0 and $t_0$ from $\mathbf{x}$, then the ray takes on the radiance at it. Otherwise, if there is an occluder from $t_0$ to $t_1$, the ray takes on the radiance at that occluder, and so on. More generally, if the segment from 0 to $t_0$ provides additional radiance $J$ and transmits $\beta$ amount of light from beyond $t_0$, we can merge the segment with the rest of the ray using the following equation:

$$L_i(\mathbf{x}, \omega) = J + \beta \cdot L_i(\mathbf{x} + t_0 \cdot \omega, \omega) \tag{6}$$

Note that this is identical to the equation for premultiplied alpha compositing [Porter and Duff 1984] using $\alpha = 1 - \beta$.

This drives our idea of what data should be stored at each cascade: For position $\mathbf{p} \in X_n^I$ and direction $\omega_p \in \Omega_n^I$, we calculate the transmittance $\beta_n(\mathbf{p}, \omega_p)$ as 0 if there is an occluder within the interval $\mathbf{p} + t \cdot \omega_p, t_{n-1} < t \leq t_n$ (where for convenience we set $t_{-1} = 0$), and 1 otherwise. Similarly, if the first hit along the interval is at distance $t_i$, the *radiance interval* is computed as $J_n(\mathbf{p}, \omega_p) = L(\mathbf{p} + t_i \cdot \omega_p, -\omega_p)$; and otherwise $\mathbf{0}$ if there is no hit. None of this cares about distances closer than $t_{n-1}$, which prevents issues such as a small occluder overlapping with a probe and shadowing all of its rays. For a more in-depth treatment of radiance intervals, see Sannikov [2023] and Osborne and Sannikov [2024].

We then have a simple algorithm for calculating the radiance: For every position $\mathbf{x}$, and direction $\omega$, we can interpolate the nearest probes and directions for each cascade level $n$ to approximate radiance and transmittance intervals $J_n$, $\beta_n$ for the ray segment from $t_{n-1} < t \leq t_n$ and merge these from back to front to compute the incident radiance $L_i(\mathbf{x}, \omega)$ using Eq. 6. We can then use many of these samples to integrate the rendering equation (Eq. 1). But while this does not require any additional ray tracing, in order to get the benefits of the angular resolution of the $n$th cascade we would need to use $|\Omega_n^I|$ directions, which is computationally infeasible. So, we can only use this procedure when the BRDF is highly anisotropic and the number of directions which contribute to the radiance is small.

For a diffuse BRDF, we need a different approach. Since the multiplier term in the rendering equation $\rho(\mathbf{x}, \omega_o, \omega_i)(\omega_i \cdot \vec{\boldsymbol{n}}(\mathbf{x}))$ is smooth, we can get high quality estimations by discretizing over a small number of sample directions, so long as the radiance stored for each direction averages the incident radiance across a wide angle rather than containing a single sample. So, we define $I_n(\mathbf{p}, \omega_p)$ for each cascade as an approximation to the average radiance across all angles closer to $\omega_p$ than any other element of $\Omega_n^I$, from distances further than $t_{n-1}$ (so that $I_0$ averages the full radiance starting from 0). We can then compute this by combining the cones for the higher cascade within the area and merging with the traced radiance interval using Eq. 6. Define $m_n(\omega)$ as the function mapping a direction $\omega$ to the nearest direction in $\Omega_n^I$, and define $m_n^{-1}(\omega_p) = \{\omega_q \in \Omega_{n+1}^I \mid m_n(\omega_q) = \omega_p\}$. Then, we have:

$$I_n(\mathbf{p}, \omega_p) = J_n(\mathbf{p}, \omega_p) + \beta_n(\mathbf{p}, \omega_p) \sum_{\omega_q \in m_n^{-1}(\omega_p)} \frac{\mathbf{Interp}(I_{n+1})(\mathbf{p}, \omega_q)}{|m_n^{-1}(\omega_p)|} \tag{7}$$

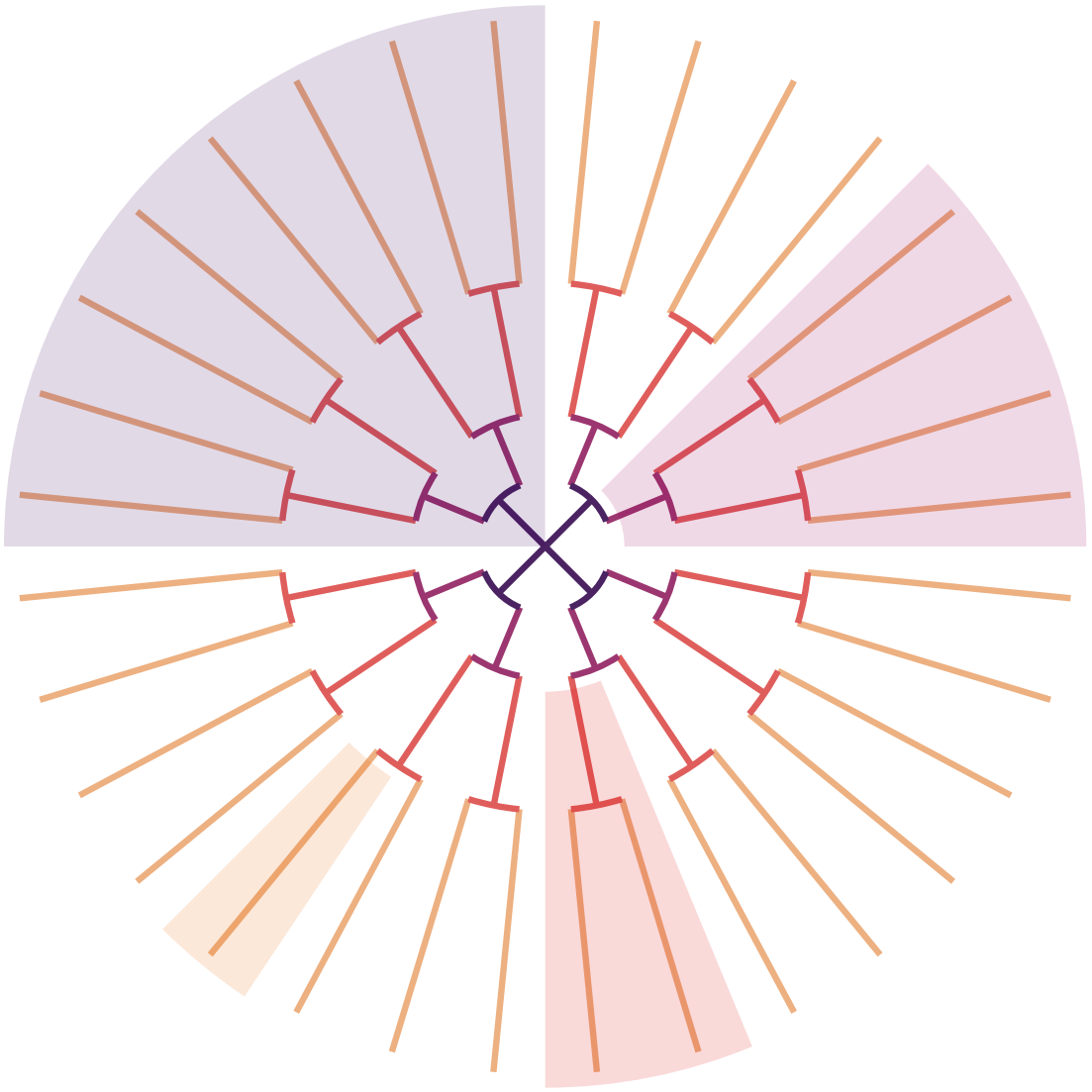

Fig. 5. The $J_n$ and $\beta_n$ intervals merged to compute radiance at a point for the probe structure in Fig. 4. One of the calculated cones $I_n$ is also shown per cascade.

We visualize this for 2D RC in Fig. 5. Finally, we calculate the radiance $L(\mathbf{x}, \omega)$ via the same approach as in normal radiance probes (Eq. 2), using the same interpolation function to estimate $I_0(\mathbf{x}, \omega_p)$ around arbitrary positions.

#### 3.2.1 *Analysis.*

The total amount of computation and raytracing performed by RC to calculate a single bounce of light across a scene is,

$$\sum_{n=0}^{N-1} \frac{V\ |\Omega_n^I|}{(\Delta_n^s)^d} = \frac{V\ |\Omega_0^I|}{(\Delta_0^s)^d} \sum_{n=0}^{N-1} \left(\frac{K}{2^d}\right)^n \tag{8}$$

where $d$ is the dimensionality of the probe structure, $V$ is the size of the volume or area we are initializing probes over, and $N$ is the number of cascades; with $l = 4$, typically only 4 or 5 cascades are necessary to have rays spanning the entire scene (with the last cascade's rays sampling the environment map upon miss).

This makes 2D implementations of RC fairly easy: We can set $K = 2$ and $l = 4$ (which due to the reduced dimensionality means $\Delta_n^\omega$ still scales the same as $\Delta_n^s$), and get high quality results with one cascade 0 (c0) probe per pixel, and 4 directions per c0 probe. We then have that $\frac{K}{2^d} = \frac{1}{2}$, so the total amount of work is bounded for an arbitrary number of cascades (so long as there is more than 1 probe per cascade). We can get higher quality results setting $K = 4$, however it is not necessary for most applications and Holographic Radiance Cascades provides better results for the same amount of computation.

In screenspace RC variants, probes are initialized at specific points on the depth buffer, using a depth-aware interpolation method instead of bilinear to avoid light leaking near depth discontinuities. This results in the probe structure still being two-dimensional. To achieve the same error scaling as 2D RC, we increase angular resolution in both axes simultaneously, which results in

$K = 4$, and then $\frac{K}{2^d} = 1$, so all cascades have the same cost.[1]

However, this scaling poses a problem for full 3D RC. Suppose we have some cascade $n$ having the same spatial and angular resolution as a normal irradiance probe scheme (so the probes would be spaced around 1 meter apart), and want to use RC to add increased details for smaller scale effects. Then, we would have to add a few cascades below $n$; but as $\frac{K}{2^d} = \frac{1}{2}$, each lower cascade becomes progressively more expensive, and so the total cost becomes dominated by the lower cascades. This is despite the fact that most of the probes in the lower cascades will be wasted, as the spacing will be fine enough that in most scenes there will be large empty areas where probes do not contribute to any surfaces (which is not the case for normal irradiance probe spacings). This brings us to our first contribution.

## 4 Sparse Radiance Cascades

Instead of allocating a large number of probes which are not used for future computation, we utilize a hashmap in order to sparsely populate the RC structure near visible surfaces only. To initialize probes near surfaces, we iterate over all pixels on screen and insert the nearest cascade 0 (c0) probe to that point in world-space. We then iterate over every c0 probe and initialize the nearest probe in c1, and so on for all cascades; we show a 2d version of this in Fig. 6. Due to how probes are spaced, this is equivalent to initializing the nearest c$n$ probe to each position visible. We do not initialize all trilinear neighbors of each probe for performance reasons, as avoiding it halves the amount of probes for little quality loss. As a result, we need to only interpolate between existing probes, which we do using a sparse trilinear approach: we add the values at the existing probes used for trilinear interpolation multiplied by their trilinear weight, and then divide by the total weight for all probes to renormalize.

We also need to calculate ray directions for the cascades. While in 2D just using evenly spaced angles for $\Omega_n^I$ is the obvious and optimal solution, in 3D there are many ways of semi-evenly distributing points on a sphere. We use the equal-area spherical mapping in Algorithm 2 to map the unit sphere of directions to $[0,1]^2$, and then generate points $\omega_{n,u,v} = \textsc{DecodeDir}\big((u+\frac{1}{2})/2\Theta_n, (v+\frac{1}{2})/\Theta_n\big)$ for some angular resolution $\Theta_n$.[2] Then, we get $\Omega_n^I = \big\{\omega_{n,u,v} \mid u \in 0..2\Theta_n - 1, v \in 0..\Theta_n - 1\big\}$. Having twice the resolution on the $x$-axis is necessary to have approximately even spacing between directions. We can easily find the nearest c$n$ direction to a cascade $n+1$ direction as $m_n(\omega_{n+1,u,v}) = \omega(n, \lfloor u/2 \rfloor, \lfloor v/2 \rfloor)$, which consistently maps four directions to one.

Merging and shading pixels then follows identically to normal RC with the adjusted interpolation method. We summarize the full algorithm in Algorithm 1.

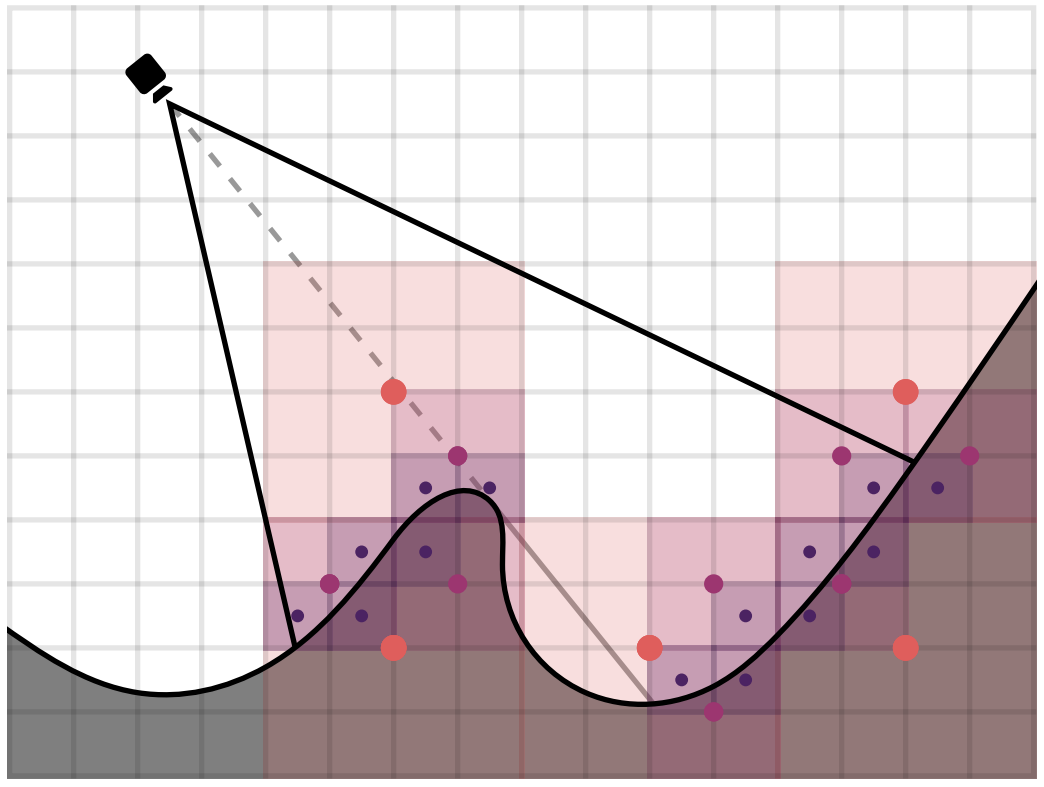

Fig. 6. Sparse probes in Radiance Cascades, with levels following the same coloring as in Fig. 4. Obscured probes are not initialized.

**Algorithm 1:** Sparse Radiance Cascades

**function** $\textsc{SparseRC}()$:
  *// Initialize probes.*
  **for** pixel $p$ on screen:
    $\mathbf{x} \leftarrow$ world position at pixel $p$
    $\textsc{Insert}_0(\textsc{Nearest}_0(\mathbf{x}))$
  **for** cascade level $n$ from 1 to $N-1$:
    **for** probe $\mathbf{p}$ in cascade $n-1$:
      $\textsc{Insert}_n(\textsc{Nearest}_k(\mathbf{p}))$
  *// Trace intervals.*
  **for** cascade level $n$ from 0 to $N-1$:
    **for** probe $\mathbf{p}$ in cascade $n$ and $\omega_{n,u,v}$ in $\Omega_n^I$:
      compute radiance interval $J_n(\mathbf{p}, \omega_{n,u,v})$
      and transmittance $\beta_n(\mathbf{p}, \omega_{n,u,v})$ using ray tracing
  *// Merge cascades.*
  **for** cascade level $n$ from $N-1$ to 0:
    **for** probe $\mathbf{p}$ in cascade $n$ and $\omega_{n,u,v}$ in $\Omega_n^I$:
      compute radiance cone $I_n(\mathbf{p}, \omega_{n,u,v})$
      from $J_n$, $\beta_n$, and $I_{n+1}$ using Eq. 6
  *// Shade pixels.*
  **for** pixel $p$ on screen:
    $\mathbf{x} \leftarrow$ world position at pixel $p$
    compute $L(\mathbf{x}, \omega_o)$ using Eq. 2
*// Gets the nearest probe in cascade $n$ to a position $\mathbf{x}$.*
**function** $\textsc{Nearest}_n(\mathbf{x})$:
  **return** $\Delta_n^s(\lfloor \mathbf{x}/\Delta_n^s \rfloor + \frac{1}{2})$
*// Inserts the probe at position $\mathbf{p}$ (assumed to be a half-integer multiple*
*// of $\Delta_n^s$) into the cascade $n$ hashmap.*
**function** $\textsc{Insert}_n(\mathbf{p})$:
  *// Code omitted; see Section 6 for details.*

### 4.1 Levels of Detail

Due to the perspective transform in (most) 3D renders, further away objects take up less screen space, and so do not need the same probe density as nearby objects to achieve an identical quality; additionally, using the same probe density would result in unbounded memory usage for very distant objects. So, we introduce a level of detail (LOD) system for our probes in the same fashion as many

[1] *Path of Exile 2*'s implementation actually uses $K = 2$, however this only functions due to tracing the depth buffer, the top-down camera view, and the vertical size/complexity of the scene not scaling with distance.

[2] Surprisingly, the octahedral [Engelhardt and Dachsbacher 2008] and Clarberg octahedral [Clarberg 2008] schemes result in worse quality despite having a more uniform distribution.

**Algorithm 2:** Equal-Area Spherical Mapping

```
function EncodeDir(ω):
    ϕ ← atan2(ω_y, ω_x)
    return (frac(ϕ/2π), (ω_z + 1)/2)
function DecodeDir(u, v):
    ϕ ← u · 2π
    z ← 2v − 1
    r ← √(1 − z²)
    return (r · cos(ϕ), r · sin(ϕ), z)
```

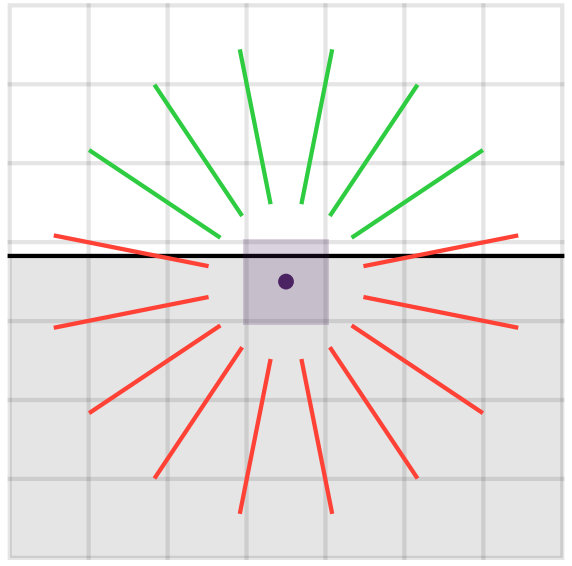

Fig. 7. An illustration of over-occlusion by 25% due to probe positioning.

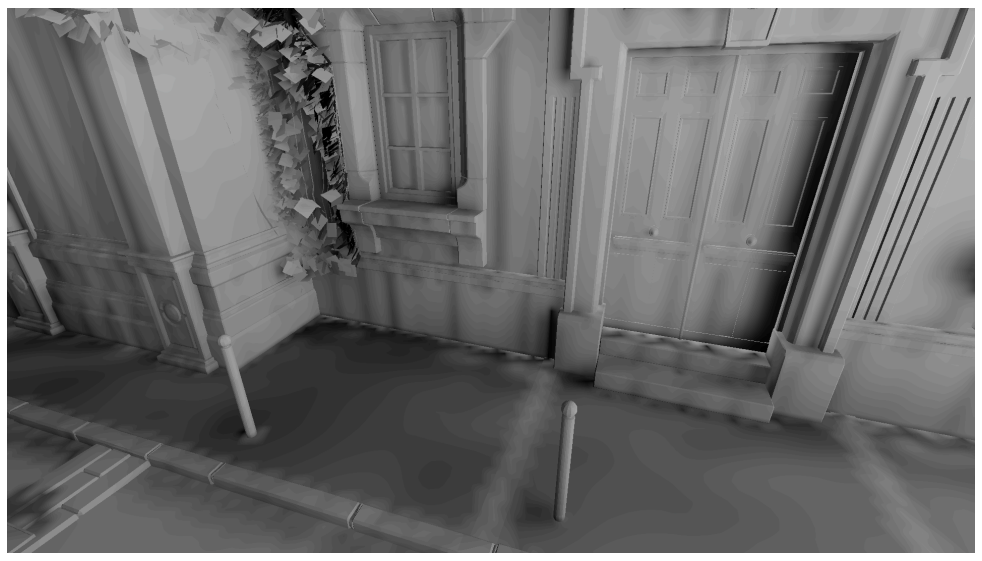

Fig. 8. Naive Sparse RC in Bistro [Lumberyard 2017]; aliasing errors are a result of probes being recessed into surfaces. c0 rays start from $t_{-1} = \Delta_0^s$.

other radiance probe algorithms.[3] Each probe stores an additional coordinate indicating its LOD, with each higher level increasing the probe spacing $\Delta_0^s$ and base ray length $t_0$ by a factor of 2. Different LODs do not interact with each other; probes of one LOD initialize the nearest higher-cascade probe of the same LOD, and when interpolating for merging and shading, we only consider probes from the same LOD. We choose the LOD for probes upon initialization based on the binary logarithm of the Chebyshev distance from the camera (it produces fewer artifacts than Euclidian due to the grid alignment of the probes), so that probes stay the same size on-screen.

To prevent the seam between LODs from being visible, we shorten the starting distance of each LOD by 0.9 to overlap with the previous LOD, and linearly blend between them across the overlapping interval when computing the final on-screen irradiance.

## 5 Ray Splitting

However, this version of Sparse Radiance Cascades still has a standard flaw of radiance probe techniques: probe positions can end up blocked within objects, which causes unnecessary darkening when interpolating. This is amplified as we intentionally spawn probes near surfaces rather than in open space. Even if we adjust c0 rays to start away from the probe position to avoid the worst of the problem, probe positions being recessed into surfaces can still cause bias due to rays facing outwards from the surface being occluded immediately, as shown in Fig. 7. The result of these errors are demonstrated in Fig. 8.

To solve this problem, we take inspiration from SHaRC [NVIDIA 2024]: we do not need to spawn rays exactly from the probe positions; instead, we can spawn rays from surfaces on-screen (so that they are not immediately occluded), and insert them into our probe structure.

Radiance Cascades uses merging to assemble multiple intervals into an infinite ray to reconstruct radiance. In *ray splitting*, we invert this by splitting rays into multiple intervals to deposit into our RC data structure. Suppose we have a ray starting at position $\mathbf{x}$ in direction $\omega$ hitting at a distance $t$, where $t_{n-1} < t \leq t_n$. Then, we know that the $n$th cascade is occluded in that direction, and all lower ones are transparent. So, for all cascades $k < n$, we can estimate $J_i(\mathbf{p}, \omega_p) = \mathbf{0}$, $\beta_i(\mathbf{p}, \omega_p) = 1$, where $\mathbf{p} = \text{Nearest}_k(\mathbf{x})$ and $\omega_p = m_k(\omega)$ are the nearest probe position and direction to the ray in that cascade. Similarly, for the $n$th cascade we can conclude that $J_n(\mathbf{p}, \omega_p) = L_i(\mathbf{x}, \omega)$ and $\beta_n(\mathbf{p}, \omega_p) = 0$.

We then deposit these values into the associated probe and direction in the hashmap. Since there may be multiple contributions, we modify the hashmap to store a cumulative sum of all radiance and transmittance values deposited into that direction, and compute the actual $J_n$ and $\beta_n$ values for merging by dividing by the number of rays deposited.

There are two ways of handling higher cascades: We can either choose to not deposit anything to them, as the ray did not hit anything within the intervals they cover, or we can extend the ray to all remaining cascades and deposit the same radiance and zero transmittance to them as well. However, the latter option results in noticeable bias due to shorter rays providing a larger contribution to the final radiance when averaged together, so we choose the former option (although it is a little noisier).

Our algorithm then proceeds as follows: After initializing probes, we spawn rays from surfaces displayed on screen, and for each ray we atomically add the radiance and transmittance contributions (and increment the count) to the nearest probe and direction in each cascade level that the ray provides information for. After all rays are deposited, we merge as normal, but ignore directions with zero count to prevent division by zero.

### 5.1 Sampling

One concern with the ray splitting algorithm is that there is no guarantee that every direction of every probe will have at least one value before merging. In fact, this is somewhat a feature: probe directions that point into surfaces will not be used in the final calculation so we avoid computing them unnecessarily by only spawning rays facing away from surfaces. Similarly, in an enclosed room all rays may terminate before reaching the $n$th cascade and we would not end up depositing anything to it; but if that is the case, the lower cascades must be fully opaque, and so we would not read anything from the $n$th cascade either.

However, we need all used directions to be filled in order to avoid noise. Conveniently, each cascade requires approximately the same number of rays to achieve this: if we consider a scene only containing a flat axis-aligned plane, then each cascade has 4x less probes than the previous, but each probe has 4x more directions

[3] In a rather unfortunate coincidence, most literature refers to this as 'cascades' (e.g. cascaded shadow maps), however to avoid confusion we will refer to them as LODs exclusively.

(assuming $K = 4$), so the total number of directions is constant per cascade. Most scenes have the same scaling due to being effectively two-dimensional, with the exception of foliage and thin rods. Additionally, due to our level of detail system, probes in the same cascade are likely to have similar projected size on-screen (so long as the camera is not seeing them obliquely) so we can ensure that they all get enough rays by spawning a constant number per pixel.

Simply using uniformly random directions results in significant inefficiencies. Higher cascades become significantly less likely to be filled in than the base cascade, as by the coupon collector's problem [Tan et al. 2025] the amount of samples required to fill $k$ buckets scales with $O(k \log k)$, and each higher cascade has exponentially more buckets per probe.

So, we use the $R_2$ low-discrepancy sequence [Roberts 2018] to generate points on the unit square, which we then map to directions using the equal-area mapping in Algorithm 2. We need that at each cascade level, all directions are semi-uniformly covered by rays, so we assign probes that share the same nearest next-cascade probe to contiguous segments of the $R_2$ sequence so that the rays for the next-cascade probe are also a contiguous segment of the sequence and so are evenly distributed. The entire sequence is then jittered to distribute errors temporally. This is done similarly to a hierarchical prefix-sum; we first count the number of rays within each probe from lowest to highest cascade, and then compute offsets for probes in reverse order. We show pseudocode for this in Algorithm 3, and demonstrate the improvement over random sampling in Fig. 9.

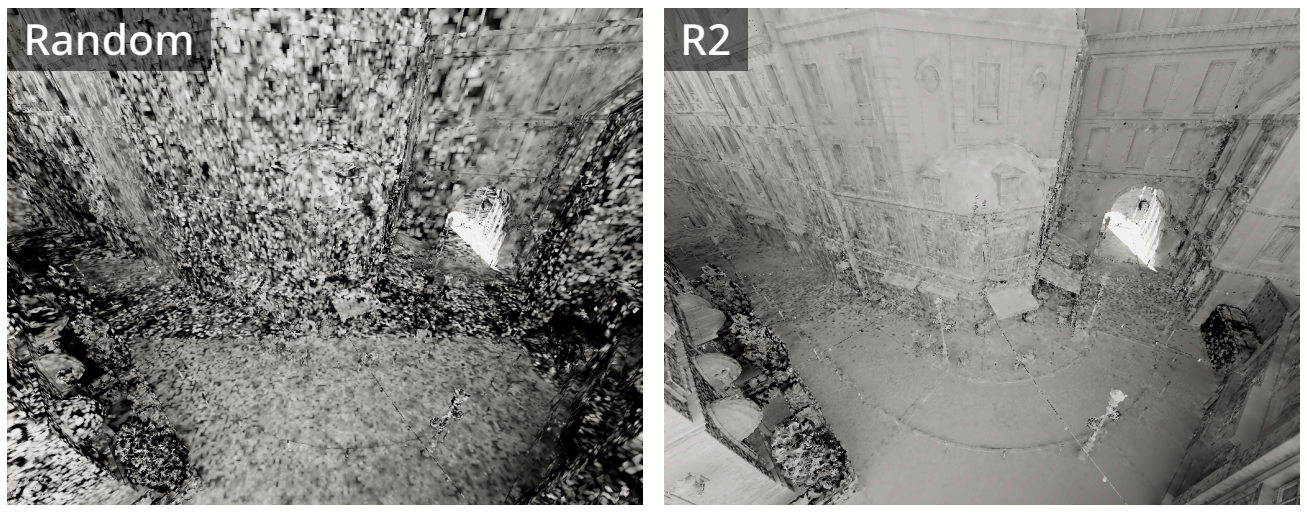

Fig. 9. Comparison of uniform random sampling and $R_2$-based sampling for a single frame of Bistro.

## 5.2 Temporal Accumulation

An additional benefit of this approach is that we can easily temporally accumulate rays: In a semi-static scene, there is no need to fully recompute probes that existed in previous frames; instead, we can blend the previous frame's cascades with the new intervals from the current frame. As our algorithm functions entirely in world-space, we do not need to worry about reprojection due to camera movement (we leave handling movable objects to future work). We can additionally use this to reduce the number of rays traced per frame by only tracing rays until we reach a certain target number; however as this is dependent on the amount of disocclusion in the last frame, for the purposes of benchmarking we will assume that every frame casts the full number of rays.

**Algorithm 3:** Ray Generation

```
// Input: A global 2d temporal noise value.
function GenerateRays(jitter):
    num_rays ← {p ↦ 0 | p a probe}
    for pixel p on screen:
        x ← world position at pixel p
        num_rays[Nearest_0(x)] += 1
    for cascade level n from 1 to N − 1:
        for probe p in cascade n − 1:
            num_rays[Nearest_n(p)] += num_rays[x]
    offset ← {}
    global_offset ← 0
    for probe p in cascade N − 1:
        offset[p] ← global_offset
        global_offset += num_rays[p]
    for cascade level n from N − 2 to 0:
        for probe p in cascade n + 1:
            local_offset ← offset[p]
            for probe q in cascade n where Nearest_{n+1}(q) = p:
                offset[q] ← local_offset
                local_offset += num_rays[q]
    for pixel p on screen:
        x ← world position at pixel p
        p ← Nearest_0(x)
        uv ← R_2(offset[p])
        offset[p] += 1
        ω ← DecodeDir(uv + jitter mod 1)
        // flip inwards-pointing directions
        ω ← ω · sign(ω · n⃗(x))
        spawn ray from x in direction ω
```

## 6 Implementation

We implement our algorithm in Rust using the LuisaCompute [Zheng et al. 2022] framework as an abstraction over CUDA. Raytracing is handled by OptiX.

*Hashmap.* We create a separate hashmap for each cascade to store radiance and transmittance. For the key, we encode the probe position as a 64-bit integer, using 18 bits for each coordinate axis and 10 for the LOD, in order to do insertion locklessly with linear probing and atomic operations. The hashmap values are very large as each contains all directions for a single probe, so we instead store an index into an indirection buffer containing the actual data. This allows us to run our hashmap at very low occupancy without consuming a large amount of memory. Additionally, we recreate our hashmaps every frame from the indirection buffer, so we can easily remove old probes without having to handle deletion in the hashmap itself.

*Merging.* We store directional data such as $J_n$, $\beta_n$, and $I_n$ in Morton order, which ensures that at each cascade, the data for all directions which are merged into the same direction in the lower cascade are stored contiguously. As a result, our merging function can do coherent reads during merging, and use warp-wide shuffles to share the data between threads to average the cones together. We additionally perform *pre-averaging*: since the only use of $I_n$ for cascades higher than 0 is to merge into the lower cascade, we directly compute the averaged value used for the next step of merging rather than storing them separately. Another optimization we use is to make each block of our merging shader responsible for only a single probe; this means that the higher-cascade probes used for interpolation are shared across the block, so we can perform lookups for them in parallel using shared memory.

*Shading.* For the final shading step, we need to compute the outgoing radiance via the rendering equation. If we do this naively by gathering all directions for the nearest probes and interpolating them for each pixel, the cost quickly becomes prohibitive: for 32 c0 directions, this results in up to 256 unfiltered texture samples per pixel. To avoid this, we first compute directional irradiance per probe by evaluating the rendering equation for each direction (assuming that the BRDF is Lambertian) and store this in a texture. Then, we sample this texture in the direction of the normal for the nearest probes and interpolate those values to get the final radiance, which is a maximum of 8 filtered texture samples per pixel. Specifically, we calculate a 6x6 octahedral-mapped [Engelhardt and Dachsbacher 2008] irradiance field, and then do an additional copying pass to add a single-texel border around the texture to handle interpolation at the edges.

*Multibounce.* We extend our algorithm to support multiple bounces by using the hit point of rays traced in the previous frame to spawn rays and probes for a secondary cache. We then sample from the secondary cache for all ray hits (which allows the secondary cache to sample itself for 'infinite' bounces via temporal accumulation). As most of the algorithm is screen-independent, we can reuse our probe initialization, ray generation, and merging code, with the only change being the input normal and positions. We do not initialize probes that require three bounces currently.

We use secondary cache probes that are two levels of detail higher than the equivalent primary cache to reduce cost, as the second bounce information is not directly displayed and so does not need to be as precise.

## 7 Results

We use length scaling $l = 4$, branching factor $K = 4$, base number of directions $|\Omega_0^I| = 32$ (using $\Theta_0 = 4$), base ray length $t_0 \approx 1.6\Delta_0^s$, and $N = 4$ cascades. We also compare to an implementation of DDGI, which we implemented following Majercik et al. [2021] with the final shading directly translated from the RTXGI-DDGI implementation [NVIDIAGameWorks 2024].

*Performance.* Our method takes approximately the same amount of time in most of our tests as probes quantity is nearly constant due to LODs. We break down the performance of each stage of our algorithm in Table 2; stage names follow Algorithm 1, with *Generate rays* referring to Algorithm 3, and *Trace rays* being the calculation of the spawned rays. We have not included the cost for creating the G-buffer or tracing shadow rays, as those steps are not specific to our algorithm. Note that these metrics are not indicative of the maximum possible performance for Split RC as our implementation is missing many micro-optimizations; for example, it is currently using 32-bit floats to store radiance.

Table 2. Performance breakdown of single-frame Split RC for Fig. 10 and Fig. 13, running on a RTX 3080 Laptop GPU. All numbers are in milliseconds.

| Stage | Sponza | San Miguel |
|---|---|---|
| Initialize probes | 1.2 | 1.3 |
| Generate rays | 0.4 | 0.6 |
| Trace rays | 1.3 | 4.5 |
| Deposit | 2.2 | 1.9 |
| Merge | 1.0 | 0.8 |
| Shade pixels | 2.5 | 2.4 |
| **Total** | **8.6** | **11.5** |

The *Trace Rays* stage is entirely dependent on raytracing performance, and so is highly variable: the San Miguel scene has complex geometry such as foliage, and so takes 3x more time to trace than Sponza.

*Quality.* We showcase Split RC in the Sponza (© Frank Meinl), San Miguel (© Guillermo M. Leal Llaguno), and Conference Room (© Anat Grynberg and Greg Ward) scenes [McGuire 2017]. We show single-frame evaluations of our algorithm, as temporal accumulation can hide performance problems and has behavior dependent on camera movement, which is hard to measure. As most other real-time global illumination algorithms utilize temporal accumulation, we also show a converged version using a smaller probe spacing and ten frames of accumulation; due to the structure of RC, there is little benefit of increasing sampling past that point as most remaining error is from interpolation. Apart from in Fig. 1 and Fig. 12, we only show a single bounce of light in our images, as secondary bounces reduce contrast and make errors harder to distinguish. DDGI results are computed using a 32x16x32 probe grid and accumulated until there is no noise (apart from the 1 frame version in Fig. 10). We note a few features of our algorithm:

- *Shadows:* Our method can capture soft penumbrae particularly well, as seen in Fig. 11; the cascade level that captures the penumbra increases with the distance to the occluder, which results in the spatial discretization of radiance matching the penumbra's width across the entire scene.
- *Ambient Occlusion:* Our method can provide medium-scale ambient occlusion effects as probes are spaced finely on-screen; this is shown clearly in the second row of cutouts for Fig. 10. DDGI does not show any ambient occlusion, which is a natural result of its probe spacing being larger than the distance from the wall to the banner.
- *Artifacts:* Split RC misses details smaller than the base probe spacing, as seen in Fig. 15, and can also exhibit light leaking artifacts from interpolation during merging, which results in brightening of shadows in Fig. 10. The small number of c0 directions can also contribute to bias, although the loss in visual quality is minimal. Additionally, Split RC has overblurring when trying to resolve hard shadows: in Fig. 13, the shadow of the plant in the cutout loses some of the details apparent in the reference.

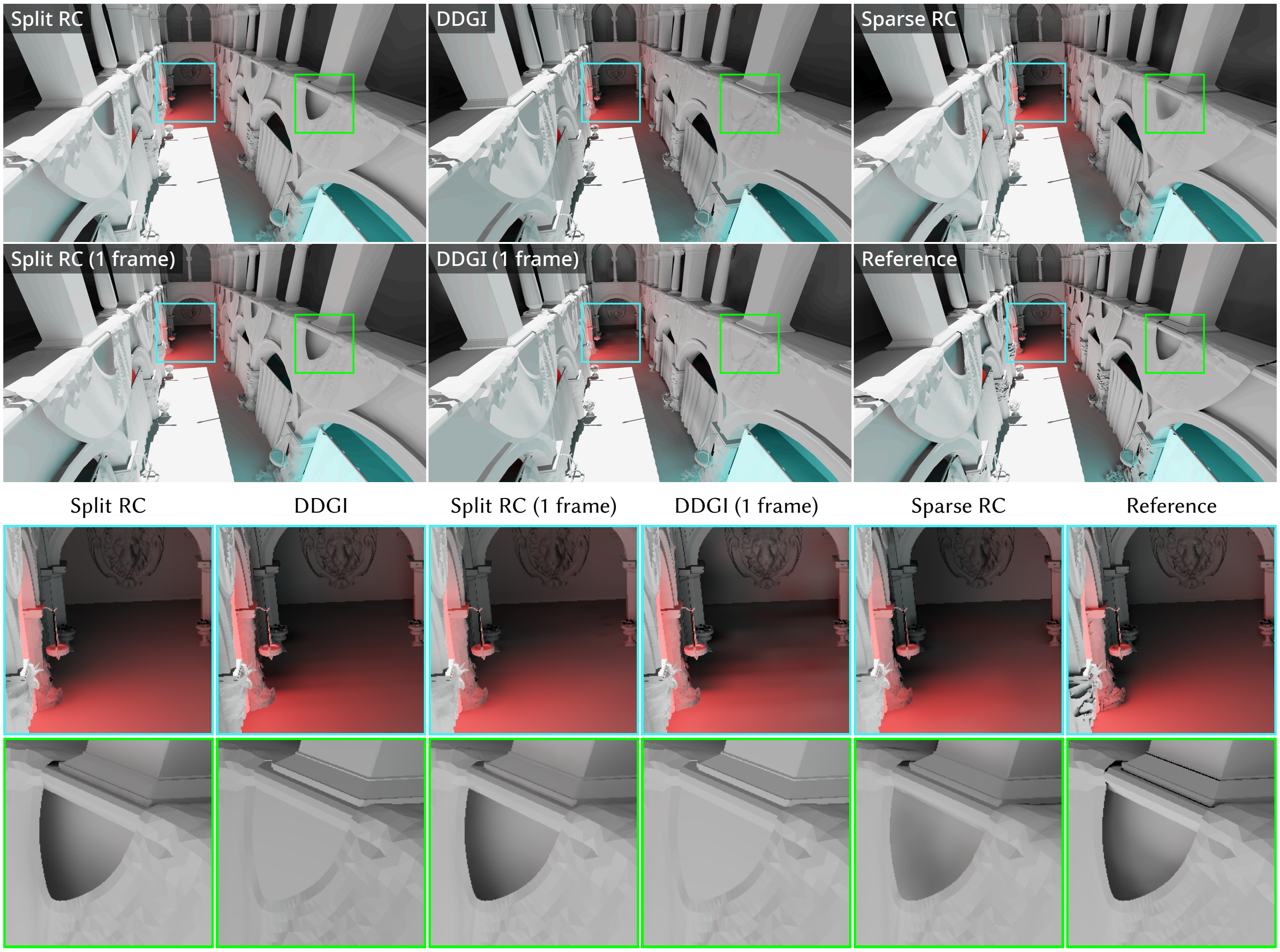

Fig. 10. Comparison of Split RC, DDGI, normal Sparse RC (single frame), and a path-traced reference for the Sponza scene. DDGI (1 frame) was computed in a single frame with an equal ray count to Split RC (1 frame); it has 1666 probes contributing to surfaces on-screen with 1245 rays each.

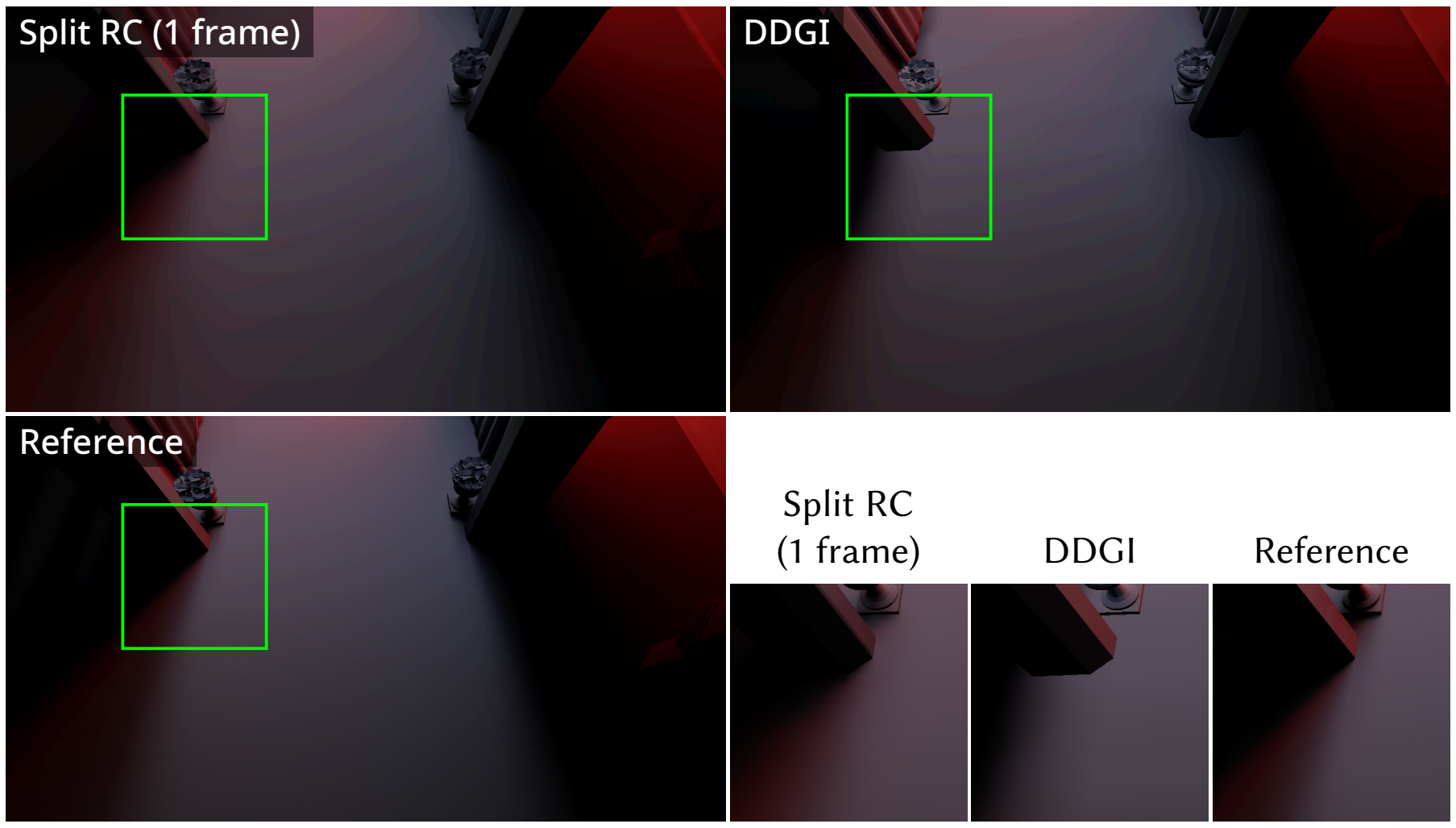

Fig. 11. Comparison of single-frame Split RC, DDGI, and a path-traced reference for Sponza near the first cutout in Fig. 10. DDGI fails to resolve the penumbra due to it being smaller than the probe spacing.

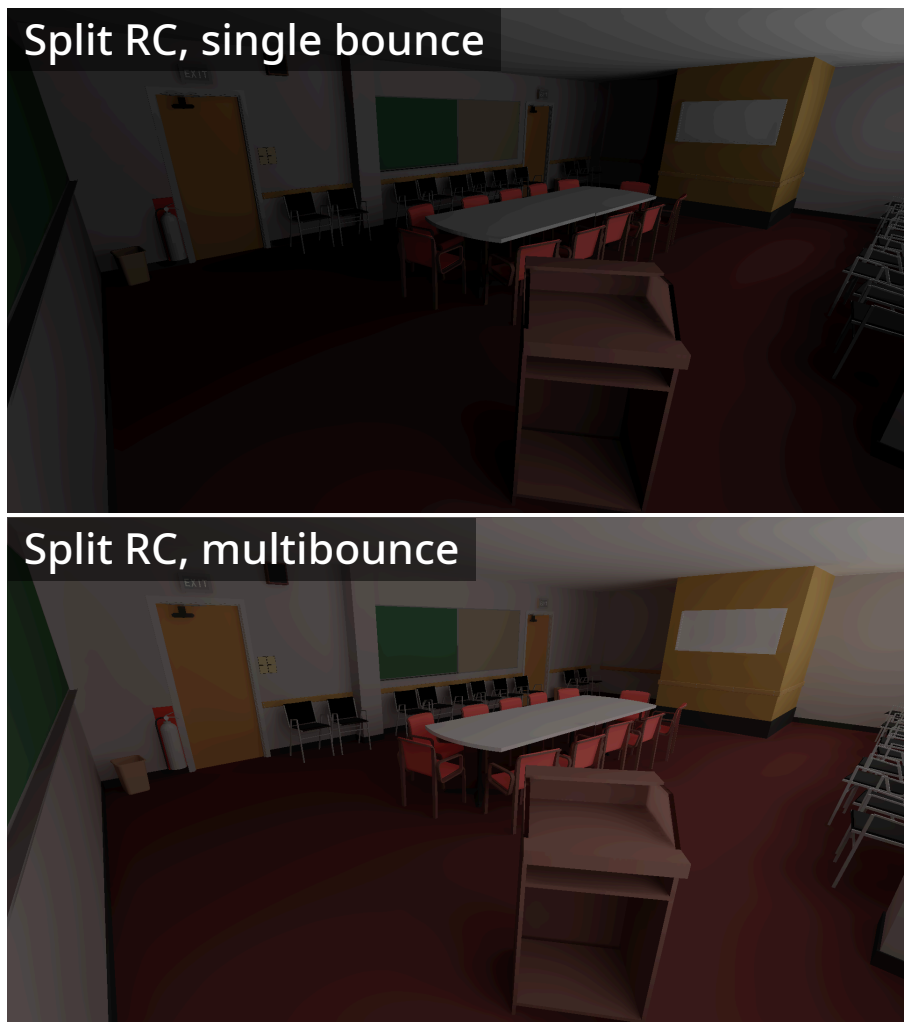

Fig. 12. Comparison between normal and multi-bounce Split RC in the Conference Room scene.

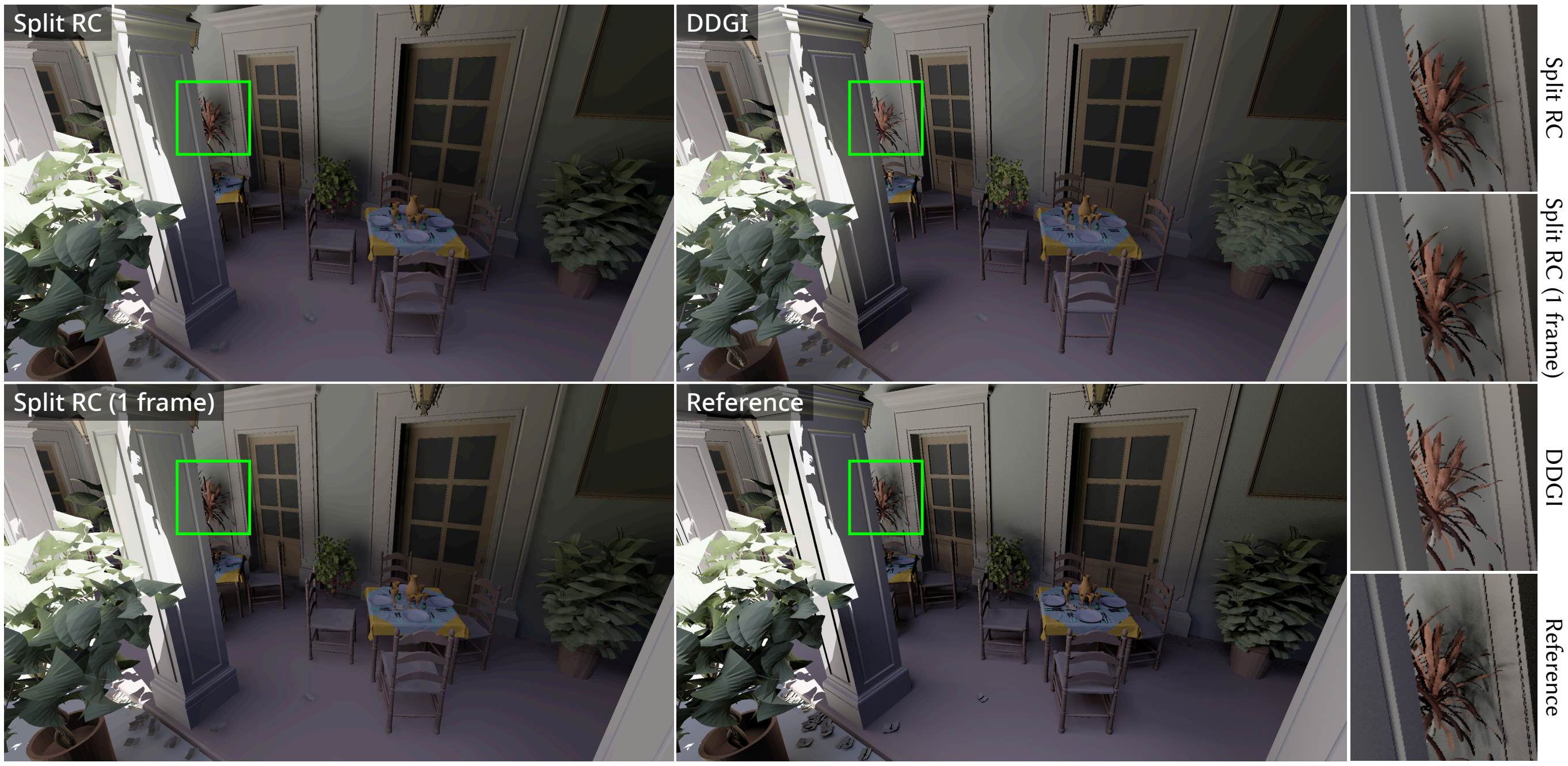

Fig. 13. Comparison of Split RC, DDGI, and a path-traced reference for the San Miguel scene. The shadows cast by the foilage are missed by DDGI due to probes being relocated away from the occluding geometry, while Split RC overblurs them.

## 7.1 Extensions

In this section, we demonstrate a few additional capabilities of our algorithm which we have not fully developed.

*Specular Effects.* As mentioned in Section 3.2, the RC data structure can be used to efficiently approximate single rays without additional tracing by combining intervals in the same direction. We can use this to handle rough specular effects by approximating it with a cone: we lookup the cascade with $I_n$ size similar to our target (for example in Fig. 14, this would be cascade 2) and merge it with $J_n$ from previous cascades to get the result. Unfortunately we cannot use RC for precise effects such as reflections as the spatial resolution loss of higher cascades becomes very apparent, so for a full rendering system this would need to be combined with a different algorithm.

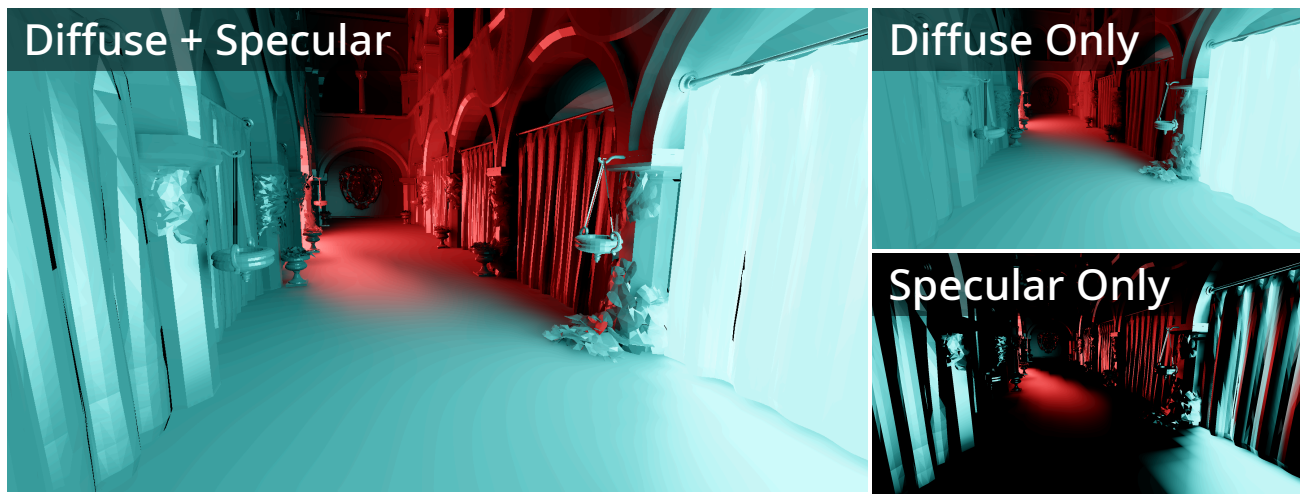

Fig. 14. Specular effects with Split RC. Notice that in the bottom right corner some spatial grid artifacts are noticeable in addition to the blurring.

*C(-1) merging.* Instead of starting c0 rays from the probe position, we can start them from a small distance $t_{-1}$ (commonly set to be $\Delta_0^s$ for normal RC as that can prevent aliasing) and merge the values with a "cascade −1", which we calculate at every pixel with screen-space raymarching similarly to Therrien et al. [2022]. This removes problems detecting features smaller than the base probe spacing; we show a comparison of this in Fig. 15. Our current unoptimized implementation uses directional radiance information for each pixel, which requires an expensive gather step rather than just sampling irradiance; this takes an additional 23ms per frame, despite the actual C(-1) calculation only being 1.4ms extra. This is avoidable by treating C(-1) as an ambient term instead and multiplying with the radiance, however that would result in the loss of directional lighting information.

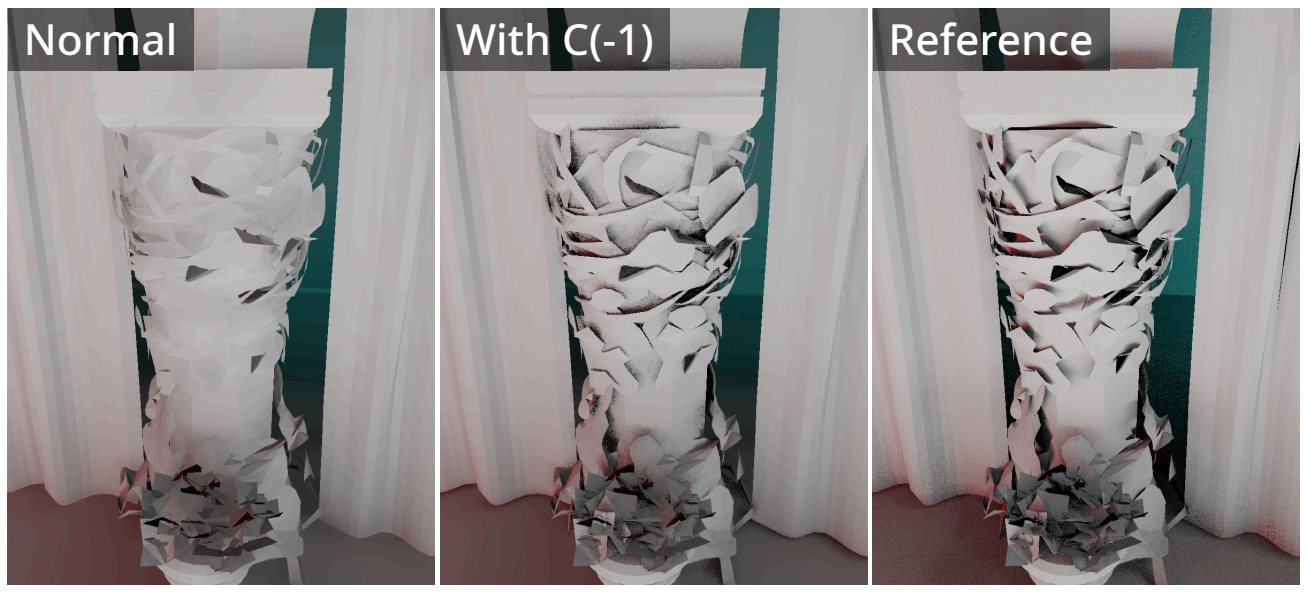

Fig. 15. Comparison of a close-up view of normal Split RC and Split RC with C(-1) merging. The leaves on the pillar have details smaller than the c0 probe spacing, resulting in normal RC being unable to capture their shadows, while C(-1) merging can capture them well.

*Denoising.* Our formulation of Split RC traces a ray from each pixel, and then uses those rays, in combination with normal, depth, and (assumed to be Lambertian) BRDF to compute a filtered image. As such, we can treat it as a denoiser that utilizes occlusion and hit depth to decide whether rays can be shared between nearby pixels. Considering it in this fashion, we can compare to other denoisers with the same input ray information. We choose OIDN [Áfra 2026] as a comparison as it is primarily designed for offline use with no temporal information, and has a simple CPU-based API, which makes integration easy. We show these results in Fig. 16.

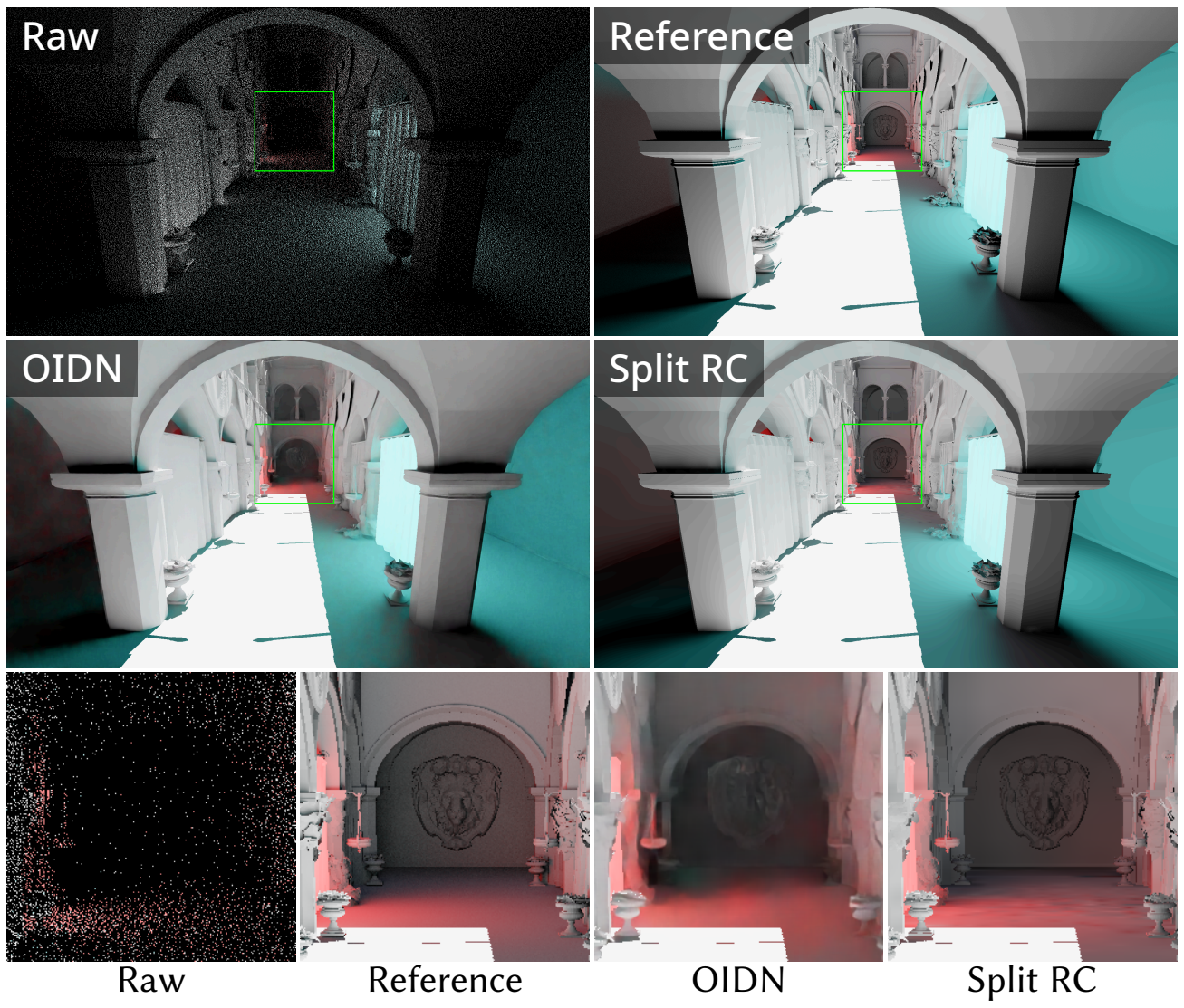

Fig. 16. Comparison of Split RC as a denoiser against OIDN. Ray directions for the raw image are computed using the sampling strategy in Section 5.1. The lion head is undersampled in the raw image, which results in significant loss of detail in OIDN, whereas Split RC is able to reconstruct it well.

Admittedly, this is a somewhat unfair comparison; OIDN is designed to take in radiance which has already been accumulated from multiple samples per pixel so it cannot use per-ray information like Split RC.

## 8 Conclusion

We have introduced a new method for 3D global illumination based on sparse world-space Radiance Cascades. Our method is capable of calculating a diffuse bounce efficiently without the need for temporal accumulation, and providing effects ranging from ambient occlusion to environment mapping within the same algorithm, while degrading gracefully for patterns too thin to accurately resolve. In addition, ray splitting avoids the standard issue of probe-based methods requiring careful positioning of probes, which is not completely ameliorated with standard sparse RC, and reduces the amount of rays required by the total number of cascades.

Ray splitting can be interpreted as an extension of path filtering techniques to filter radiance intervals rather than full rays, with longer intervals using a larger filter size and being occluded by shorter intervals.

We also demonstrate a new argument for the effectiveness of Radiance Cascades: that the errors induced by spatial and angular discretization should be simultaneously minimized, which requires multiple probe levels due to the angular error scaling with ray length. This argument drives our choice of scaling factor.

### 8.1 Limitations and Future Work

As RC captures the radiance field up to a certain frequency, it is incapable of resolving certain phenomena such as sharp mirror reflections. Additionally, Split RC can exhibit light leaking and other artifacts, particularly for geometry smaller than the base probe spacing, and is inherently biased due to use of interpolated probes.

The primary difficulty with our ray splitting scheme is that it requires finding rays that are evenly distributed across all cascades. While our current approach achieves this goal better than using white or blue noise distributed directions, there is still plenty of room for improvement; spawning rays from pixels results in up to a 4x ray quantity difference between the nearest and furthest probes within the same LOD, so approaches using a fixed ray budget per probe are worth investigating.

Our temporal accumulation method is very naive; a better approach would be to allocate more rays towards newly spawned probes (although this re-raises the question of how to evenly distribute rays), and reduce the amount of rays for older probes unless there is a change in lighting, with the aid of a multi-scale mean estimator or other methods of adaptive averaging.

Finally, it may be possible to represent Split RC entirely as a denoiser without hashing or additional storage, which could allow it to be used as a drop-in replacement for existing denoisers over low sample count path tracers while retaining the benefits of RC.

## A 2x vs 4x Length Scaling

In this appendix, we compare the quality of Radiance Cascades with different length scaling factors. For clarity, we show results in 2D, and use branching factor $K = 2$ to replicate the same angular resolution changes across cascades as $K = 4$ has in 3D. We show these results in Fig. 17. As RC functions independently of scene complexity, this simple test case is sufficient to demonstrate the performance of both scaling factors in the general case.

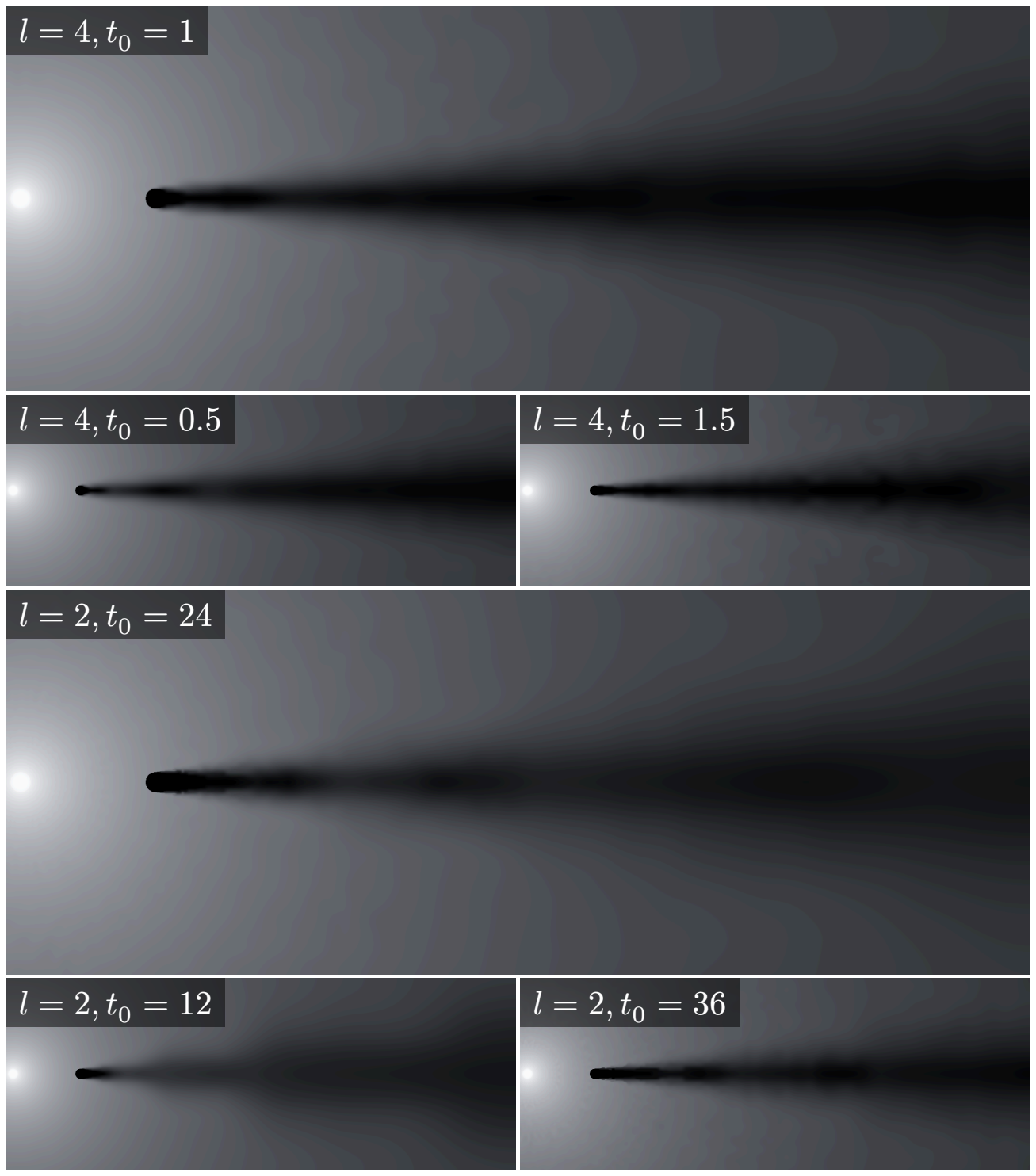

Fig. 17. Comparison of different length scaling factors for RC for a single light source and occluder. All images use 32 directions per c0 probe with one c0 probe per pixel (using units $\Delta_0^s = 1$), and are taken at a resolution of 2048x768. We chose c0 ray length $t_0$ for the large images so that aliasing is barely apparent.

In the $l = 4$ examples, aliasing (insufficient angular resolution) and blurring (insufficient spatial resolution) appear consistently across the image; in $t_0 = 1.5$, aliasing is the only error apparent while in $t_0 = 0.5$ the same is true for blurring, and in $t_0 = 1$ they are balanced. In contrast, for $l = 2$, the error gradually changes from aliasing to blurring as the distance from the light increases; all three examples have blurring at the rightmost part of the image and even the $t_0 = 12$ example has an overly-sharp penumbra right next to the occluder.

Curiously, the optimal result for $l = 4$ being $t_0 = 1.0$ disagrees with our prediction that the correct term to optimize is $\min(\Delta^s, \Delta^\omega)$; with 32 directions, the c0 ray length should be around $\frac{32}{2\pi}\Delta_0^s \approx 5.09\Delta_0^s$ to have $\Delta^s = \Delta^\omega$ instead. We believe this is due to the aliasing from angular error being more noticeable than blurring; in our actual Split RC implementation, we usually have multiple rays per direction, which reduces the aliasing and makes a longer ray length be optimal.